\documentstyle[amssymb,aps,subeqnarray]{revtex}
\input epsf
\def\csch{\mbox{csch}}
\def\coth{\mbox{coth}}
\def\Cc{C_C}    
\def\Ca{C_A}
\def\Cn{C_N}
\def\dphis{\Delta\phi_S}   
\def\dphid{\Delta\phi_D}   
\def\dphii{\Delta\phi_i}   
\def\Cs{C_S}    
\def\epss{\epsilon_S}   
\def\epsb{\epsilon_b}   

\begin{document}
\title{Asymptotic Analysis of Diffuse-Layer Effects on Time-Dependent
Interfacial Kinetics}
\author{Antoine Bonnefont and Fran\c{c}oise Argoul \\
{\small {\it Centre de Recherche Paul Pascal, Avenue Schweitzer, 33600
Pessac, France}}
\\ Martin Z. Bazant \\
{\small {\it Department of Mathematics, Massachusetts Institute of
Technology, Cambridge, MA 02139}} \\
}
\date{\today}
\maketitle

\begin{abstract}
We investigate the subtle effects of diffuse charge on interfacial
kinetics by solving the governing equations for ion transport
(Nernst-Planck) with realistic boundary conditions representing
reaction kinetics (Butler-Volmer) and compact-layer capacitance
(Stern) in the asymptotic limit $\epsilon = \lambda_D/L \rightarrow
0$, where $\lambda_D$ is the Debye screening length and $L$ is the
distance between the working and counter electrodes.  
Using the methods of singular perturbation theory, we derive the
leading-order steady-state response to a nonzero applied current in
the case of the oxidation of a neutral species into cations, without
any supporting electrolyte. In certain parameter regimes, the theory
predicts a reaction-limited current smaller than the classical
diffusion-limited current.  We also analyze the impedance of the
electrochemical cell when a small AC current modulation is added to an
applied DC current.  At sufficiently high AC frequencies, the Maxwell
displacement current is found to exceed the Faradaic conduction
current, and experimentally observed ``negative impedances'' (out of
phase AC voltage responses) are predicted close to the
reaction\--limited current.  Overall, we demonstrate that the dynamics
of diffuse charge plays a fundamental role in nonequilibrium surface
reactions when the transport of one of the reacting species is coupled
to the total interfacial reponse of the compact and diffuse layers.
\end{abstract}


\section{Introduction}

The influence of diffuse-layer structure on electrochemical surface
reactions has been studied for many years since the pioneering work of
Frumkin \cite{Frumkin:55}. The double layer is typically modeled as a
stationary outer ``diffuse layer'' (or ``Debye layer'') of charged
species in transport equilibrium and a inner ``compact layer'' (or
``Stern layer'') where the electroactive species can react at an
effective electrostatic potential $\Phi_r$ (reaction plane) which is
different from the electrode potential $\Phi_e$ by an amount $\Delta
\Phi_S = \Phi_e - \Phi_r$, the Stern-layer voltage. In such models the
only effect of the diffuse layer on interfacial kinetics is to modify
the activation energy and effectively rescale the interfacial
concentration of the reacting species
~\cite{Grahame:47,Parsons:54,Delahay:65,Delahay:66,Newman:65,Newman:66,Levie:67,MacDonald:70,Norton:90,Levine:71,Fawcett:72,Damaskin:93,Damaskin:93b,Fawcett:98}.
Within this theoretical approach, two types of models have been used:
Either (i) the diffuse layer is treated as a continuum retaining the
essential features of the mean concentration and electric field
profiles
~\cite{Grahame:47,Parsons:54,Delahay:65,Delahay:66,Newman:65,Newman:66,Levie:67,MacDonald:70,Norton:90}
or (ii) the interfacial double layer is treated as an idealized set of
nested layers (e.g. the inner and outer Helmholtz planes) consisting
of discrete molecular and/or ionic
species~\cite{Levine:71,Fawcett:72,Damaskin:93,Damaskin:93b,Fawcett:98,Fawcett:73}.
When diffuse-charge effects are incorporated into either type of
model, the diffuse layer is assumed to have a stationary, equilibrium
distribution of charges. The
dynamical effect of the diffuse layer on charge transfer kinetics has
only rarely been considered since it was first mentioned by Levich
fourty years ago \cite{Levich:49,Gierst:61}.

The present study is motivated by the idea that the dynamical charging
of the diffuse layer could also have an important effect on surface
reaction kinetics in weak electrolytes (whose Debye screening length
significantly exceeds the effective width of the compact layer). In
this work, we adopt the classical Nernst-Planck ``mean-field''
continuum model for ion transport, without considering the possible
steric effects due to finite ion sizes~\cite{Borukhov:97}. We do,
however, allow for the compact Stern layer to have a different
permittivity $\epss$ from the bulk solvent $\epsb$ as described in
Ref.~\cite{Bazant:00}, and we also consider nonlinear reaction
kinetics given by the Butler-Volmer equation without any
approximations (e.g. linearization).

Unlike previous studies, we focus here on dynamical, non\--equilibrium
effects of the diffuse-layer charging on the effective kinetic
parameters measured by impedance spectroscopy. There have been few
attempts in the literature aimed at characterizing the dynamics of the
diffuse layer and its interplay with surface reactions, and existing
studies have been mostly numerical in various
approximations~\cite{Norton:90,Levich:49,Brumleve:78,Murphy:92}. Here
we illustrate the power of asymptotic analysis to provide analytical
predictions of the statics and dynamics of realistic interfacial
double layers in response to both DC and AC currents.  Asymptotic
analysis has rarely been applied to electrochemical systems, and
previous studies have only focused on static cases with oversimplified
boundary
conditions~\cite{Newman:65,Newman:66,Grafov:62,Chernenko:63,Rubinstein:79,Henry:89}.
Recently, more realistic boundary conditions have been proposed to
model the interplay between the diffuse and compact layers, and
asymptotic analysis has provided insight into the static response of a
binary cell near the classical diffusion-limited and reaction-limited
currents~\cite{Bazant:00}. In this work, the same realistic equations
and boundary conditions are used to model the AC+DC response of a
three-species electrochemical cell (composed of a binary electrolyte
plus a neutral solute molecule). To our knowledge, this is the first
asymptotic analysis of a time-dependent, nonequilibrium
electrochemical system.

The article is organized as follows. In section~\ref{sec:eqns}, we
describe the electrochemical system and present the model equations
for ion transport and surface reactions. In section~\ref{sec:static}
leading-order asymptotic approximations of steady solutions to these
equations are derived in the case of a DC applied current. Finally, in
section~\ref{sec:imp} the asymptotic analysis is generalized to a
time-dependent case by deriving the linear response to a small AC
modulation added to a nonzero DC current.  The full details of the
time-dependent analysis will be presented elsewhere, but here we focus
on the prediction of the negative impedance which has been observed in
experiments and invoked as interpretation of the occurrence of
oscillatory instabilities in electrochemical
systems~\cite{Koper:93b,Koper:96,Strasser:99,Krischer:99}.

\section{Description of the Model Equations}
\label{sec:eqns}

\subsection{Transport Equations}

For simplicity, we consider an electrochemical cell with a planar
geometry, possessing one electrode at $X=0$ (working electrode) 
and another at $X=L$ (counter electrode), and
hence we ignore the possible effect of convection on ion
transport. (Note that we use upper-case letters to denote quantitites
with dimensions to distinguish them from the corresponding
dimensionless quantities introduced below, which are denoted by
lower-case letters.)  In order to limit our study to a small number of
species, we analyze the case of a cation C which reacts at the
electrodes to produce a neutral molecule N according to the
electrochemical reaction
\begin{equation}
\mbox{C} + z e^- \rightleftarrows \mbox{N}
\label{eq:reaction}
\end{equation}
in the presence of an anion A, which aids in charge neutralization.
For analytical convenience, we also assume that the electrolyte is
symmetric, $z_C = -z_A = z$.

The continuum fields representing the state of the solution are the
cation, anion and neutral molecule concentrations (moles/volume)
$\Cc(X,t)$, $\Ca(X,t)$ and $\Cn(X,t)$, respectively, and the
electrostatic potential $\Phi(X,t)$. (Note that we use a lower case
$t$ for dimensional time, to distinguish it from the absolute temperature
$T$.)  Although we allow for time-dependent electric fields and
currents, it can be shown that magnetic effects are negligible at the
time scales under consideration so the electric field is given by $E =
- \partial_X \Phi$. In the dilute solution approximation, the average
concentrations evolve according to the Nernst-Planck equations
describing diffusion and electromigration
\begin{subeqnarray}
\partial_{t} \Cc & = & D \partial_X^2 \Cc
  + \mu \partial_X (\Cc \partial_X \Phi ) \\ 
\partial_{t} \Ca & = & D \partial_X^2 \Ca
  -  \mu \partial_X (\Ca \partial_X \Phi ) \\ 
\partial_{t} \Cn & = & D \partial_X^2 \Cn
\label{transport_dim}
\end{subeqnarray}
where the diffusion coefficients and electrostatic mobilities of all
species are assumed to be constant and equal to $D$ and $\mu$,
respectively.  The electrostatic potential is given by Poisson's equation
\begin{equation}
-\epsb \partial^2_X \Phi = zF(\Cc - \Ca) 
\end{equation}
where $\epsb$ is the permittivity of the bulk solvent (assumed to be
constant outside the compact layer) and $F$ is the Faraday constant.

The Faradaic current density due to the electromigration (conduction)
of charged species is given by
\begin{equation}
J_F = - z F \left[ D \partial_X (\Cc - \Ca) + \mu (\Cc + \Ca)
\partial_X \Phi \right].
\end{equation}
Although we consider time-varying electric fields which are too small
to generate significant magnetic fields, we will show that they can be
large enough to produce a non-negligible Maxwell displacement
current. Therefore, the total current density is given by
\begin{equation}
J(t) = J_F(X,t) - \epsb \partial_t \partial_X \Phi(X,t)
\label{current_dim}
\end{equation}
which is the sum of the conduction and displacement current densities.
Note that $J$ is uniform across the cell since $\partial_X J =0$, and
therefore $J$ is the experimentally controlled (or measured) 
current in a time-dependent situation, not $J_F$.

\subsection{Boundary Conditions}

At each electrode, $X=0$ and $X=L$, there are three boundary
conditions on the ionic fluxes expressing mass conservation in the
electrochemical surface reaction (\ref{eq:reaction})
\begin{subeqnarray}
D\partial_X \Cc + \mu \Cc \partial_X \Phi & = & - J_F/zF \\
D\partial_X \Ca - \mu \Ca \partial_X \Phi & = & 0 \\
D\partial_X \Cn & = & J_F/zF
\end{subeqnarray}
(We adopt the sign convention that a cathodic current is negative.)
In the mean-field approximation, the Faradaic current $J_F$ at each
electrode (which contributes to the surface reaction, unlike the
displacement current) is related to the local concentrations and
potential through the Butler-Volmer kinetic equation (applied at $X=0$
and $X=L$)
\begin{equation}
J_F = -K_O \Cc \exp \left( -\alpha_O zF \Delta \Phi_S/RT \right) + K_R \Cn 
\exp \left( \alpha_R zF \Delta \Phi_S / RT \right)
\label{eq:BV}
\end{equation}
where $K_O$ and $K_R$ are the oxidation and reduction kinetic rate
contants, respectively, $\Cc$ and $\Cn$ are the interfacial
concentrations, $\alpha_O$ and $\alpha_R$ are the
transfer coefficients (which are set equal to $1/2$ below) and $\Delta
\Phi_S = \Phi_e - \Phi_r$ is the voltage drop across Stern's compact
layer mentioned above. Following Frumkin \cite{Frumkin:55}, we imagine
applying (\ref{eq:BV}) at the outer Helmholtz plane, or any other
convenient molecular distance which acts as the edge of the continuum
region. Note we do not apply the Butler-Volmer equation across the
entire interface (including both the compact and diffuse layers) as is
frequently done.

No other boundary conditions are typically mentioned in
electrochemistry textbooks because none are needed when the common
assumption of electroneutrality is made~\cite{Newman:91}. In this
work, however, since we treat diffuse charge explicitly we need
another boundary condition on the potential. Ignorance of this
boundary condition is typically hidden in the ``zeta-potential''
$\zeta$ (the potential drop across the diffuse layer
$\Delta \Phi_D$) or equivalently the total charge in the diffuse
layer. In colloidal science these quantities can be assumed to be
constant properties of a surface, but in electrochemistry, the zeta
potential of an electrode should be determined self-consistently from
the microscopic electrochemical boundary conditions. We will see that
the zeta potential of a working electrode can vary widely with
electrochemical conditions, especially in time-dependent situations.

A general expression for the missing boundary condition has recently
been proposed based on the Grahame model of the interface, which
contains the Stern model as a special case~\cite{Bazant:00}. Here we
adopt the Stern model, which postulates a constant capacitance $\Cs$
for the compact layer
\begin{equation}
\Delta \Phi_S = \left\{
\begin{array}{ll} 
 - \epss \partial_X \Phi / \Cs & \ \mbox{ at } X=0 \\
 \epss \partial_X \Phi / \Cs  & \ \mbox{ at } X=L
\end{array} \right.
\end{equation}
where $\epss$ is an effective permittivity for the Stern layer.
Note that we do not consider, in this model, any chemical adsorption effect.
In this work, we model experiments in
which the current $J(t)$ is prescribed and the voltage difference
between the working electrode and the reference electrode 
$\Delta \Phi_{tot} = \Phi_{e} - \Phi_{ref}$ is
determined by solving the equations. Note that
\begin{equation}
\lambda_S = \epss / \Cs
\end{equation}
defines an effective width for the Stern layer. We do not claim that
$\lambda_S$ corresponds to any well-defined molecular distance, but
only that it captures the combined effect of the compact layer
capacitance and permittivity.  In our numerical calculations below, we
assume that the permittivity of the Stern layer is $\epss =
10\epsilon_o$, compared with the value $\epsb = 80\epsilon_o$ for a
water solvent, consistent with previous studies \cite{Guidelli:92}.
From the equation $\lambda_{S} = \epsilon_{S}/C_{S}$ and the
literature data on $C_{S} \sim 80 \mu$F.cm$^{-2}$ \cite{Guidelli:92}
an estimate of $\lambda_{S} \sim 1 \AA$ can be computed. We also
assume that $\lambda_{S}$ is independent of the local ionic
concentrations.

\subsection{Dimensionless Equations and Boundary Conditions}

The first step in any asymptotic analysis is to scale all quantities
appropriately and identify the relevant dimensionless groups.  Here we
scale length to the electrode separation $L$, time to the diffusion
time across the cell $L^2/D$, potential to the thermal voltage
$zFD/RT$, concentrations to the mean anion concentration $C^*$
\begin{equation}
C^* = \frac{1}{L} \int_0^L \Ca(X) dX
\end{equation}
and current to the classical diffusion-limited current  of Nernst
\begin{equation}
J_{DL} = \frac{4 z F D C^*}{L}.
\end{equation} 
The dimensionless variables at these scales are
\begin{equation}
x = X/L, \ \ \tau = tD/L^2, \ \ j = J/J_L, \ \ c_i = C_i/C^* \
(i=\mbox{C,A,N}), \ \ \phi = zF \Phi/RT.
\end{equation}
We also introduce the mean concentration of charged species $c(x,t)$
and the charge density $\rho(x,t)$
\begin{subeqnarray}
c &=& \frac{1}{2}(c_C + c_A) \\
\rho &=& \frac{1}{2}(c_C - c_A) 
\end{subeqnarray}
In terms of these variables, the Nernst-Planck and Poisson equations
take the dimensionless forms (using the Einstein relation $\mu = zFD/RT$)
\begin{subeqnarray}
\partial_\tau c &=& \partial^2_x c + \partial_x(\rho \partial_x \phi) \\
\partial_\tau \rho &=& \partial^2_x \rho + \partial_x(c \partial_x
\phi) \slabel{eq:drho} \\
\partial_\tau c_N &=& \partial^2_x c_N \\
-\epsilon^2 \partial^2_x \phi &=& \rho
\label{eq:eqns}
\end{subeqnarray}
where the dimensionless group $\epsilon = \lambda_D/L$ indicates that
the natural scale for charge screening is the Debye length $\lambda_D$
given by
\begin{equation}
\lambda_D^2 = \frac{\epsb RT}{2 F^2 C^*}
\end{equation}
rather than the electrode separation $L$.  In this work we consider
the typical case in which $\epsilon$ is much smaller than unity (for a
cell of typical length $L=1cm$ and $\lambda_{D}<100$nm, $\epsilon <
10^{-5}$), which is the basis for our asymptotic analysis. Since
$\epsilon$ multiplies the highest derivative in the equations, it is a
singular perturbation which must be treated using the methods of
boundary layer theory~\cite{Bender:78}. Note that the expression for
the dimensionless current (which is uniform across the cell since
$\partial j/\partial x=0$)
\begin{equation}
j(t) = j_F(x,t) + j_d(x,t)
\end{equation}
where
\begin{subeqnarray}
j_F &=& -\frac{1}{2}\left( \partial_x \rho + c \partial_x \phi \right) \\
j_d &=&  - \frac{1}{2} \epsilon^2 \partial_\tau \partial_x \phi(x,t)
\end{subeqnarray}
involves the perturbation parameter multiplying the displacement
current density $j_d$, which indicates that it only becomes important
at high frequencies (and/or small length scales).

The flux boundary conditions at $x=0$ and $x=1$ take the dimensionless
forms
\begin{subeqnarray}
\partial_x c + \rho \partial_x \phi & = & - 2 j_F \\
\partial_x \rho + c \partial_x \phi &=& -2 j_F \\
\partial_x c_N & =& 4 j_F
\label{eq:fluxbc} 
\end{subeqnarray}
where the dimensionless Faradaic current at each electrode is given by
\begin{equation}
j_F = -k_O \left( c + \rho \right) \exp \left( - \alpha_O \dphis
\right) +k_R c_{N} \exp \left( \alpha_R \dphis
\right)  \label{butler_adim}
\end{equation}
where $k_O = K_O L/4D$ and $k_R = K_R L /4D $ are the dimensionless
kinetic constants and $\dphis = zF\Delta \Phi_S /RT$. The
dimensionless Stern boundary conditions
\begin{equation}
\dphis = \left\{ \begin{array}{ll} 
- \delta \epsilon \partial_x \phi & \ \mbox{ at } x=0 \\
\delta \epsilon \partial_x \phi & \ \mbox{ at } x=1
\end{array} \right.
\label{eq:sternbc}
\end{equation}
involve a second
dimensionless group $\delta = \lambda_S/\lambda_D$, which is the ratio
of the effective widths of the compact and diffuse parts of the
interfacial double layer. The Gouy-Chapman model (with no compact
layer) corresponds to the limit $\delta = 0$ while the Helmholtz model
(with no diffuse layer) corresponds to the limit $\delta \rightarrow
\infty$. In this work, we consider the full range of values for
$\delta$, but always require $\epsilon \ll 1$.

It is ubiquitous in electrochemistry to assume that the total charge
in the cell
\begin{equation}
q(t) = \int_0^1 \rho(x,t)dx
\end{equation}
is identically zero, but this ``macroscopic intuition'' (which is
valid in the limit $\epsilon \rightarrow 0$) is flawed whenever
diffuse charge is being treated explicitly ($\epsilon>0$), as we do
here, because there is a total charge (of order $\epsilon$) in the
diffuse layers which exists to satisfy the nonlinear and asymmetric
reaction conditions at the electrodes~\cite{Bazant:00}.  

The fact that $q(t)$ can vary in time does not contradict the
conservation of law for charge because $dq/dt$ is related by
(\ref{eq:drho}) to the Faradaic currents at the electrodes
\begin{equation}
\partial_\tau q = \int_0^1 \partial_\tau \rho(x,t) dx = 2
\left[ j_F(0,\tau) - j_F(1,\tau) \right]
\end{equation}
rather than the total currents (which are equal). Since the total
current $j = j_F + j_d$ is uniform across the cell, any changes in the
total charge of the cell are caused by temporary differences between
the displacement currents at the two electrodes.

\section{Steady State Response to a DC Current}
\label{sec:static}

\subsection{The Smyrl-Newman Equation}

In this section, we derive leading-order steady-state solutions of
(\ref{eq:eqns}) and (\ref{eq:fluxbc})--(\ref{eq:sternbc}) in response
to a constant applied DC current density $j$ , which are valid after
transients have died away, following the analysis of
Ref.~\cite{Bazant:00}. Integrating the steady Nernst-Planck equations
once, we arrive at a system of ordinary differential equations
\begin{subeqnarray}
\partial_x c + \rho \partial_x \phi & =& -2j
\slabel{eq:ceq} \\
\partial_x \rho + c \partial_x \phi & =& -2j
\slabel{eq:rhoeq} \\
\partial_x c_N & =& 4j
\slabel{eq:cneq} \\
-\epsilon^2 \partial^2_x \phi &=& \rho
\slabel{eq:poisson}
\end{subeqnarray}
By subtracting the first two equations we obtain
\begin{equation}
\partial_x (c-\rho) = (c-\rho) \partial_x \phi
\end{equation}
which is easily integrated to show that (in steady-state) the
anions are in Boltzmann thermal equilibrium throughout the cell
\begin{equation}
c_A = c-\rho \  \propto \ \exp(\phi) = \exp(- z_A F\Phi/RT)
\end{equation}
since they do not react at the electrodes.

Our analysis begins by combining the three coupled equations
(\ref{eq:ceq}), (\ref{eq:rhoeq}) and (\ref{eq:poisson}) into a single
equation for the potential, which is decoupled from the fourth
equation (\ref{eq:cneq}) for the neutral species concentration (except
through the boundary conditions). Substituting (\ref{eq:poisson}) into
(\ref{eq:ceq}) and integrating we obtain
\begin{equation}
c(x) = c_o -2jx + \frac{\epsilon^2}{2} \left( \partial_x \phi
\right)^2
\label{eq:c}
\end{equation}
where $c_o$ is an integration constant to be determined by the
boundary conditions. If we substitute this expression and
(\ref{eq:poisson}) into (\ref{eq:rhoeq}) we obtain a third order
nonlinear equation for the potential~\cite{Bazant:00}
\begin{equation}
\epsilon^2 \left[ \partial^3_x \phi - \frac{1}{2}  \left(
    \partial_x \phi \right)^3 \right] + \left(2jx - c_o \right)
    \partial_x \phi = 2j 
\label{eq:martin_eq}
\end{equation}
A similar equation was first derived by Smyrl and Newman for the
electric field of a rotating disk electrode near the diffusion-limited
current~\cite{Newman:66}. Once this equation is solved for the
potential $\phi$, the concentration $c$ and charge density $\rho$ are
computed from (\ref{eq:c}) and (\ref{eq:poisson}), respectively.

The Smyrl-Newman equation (\ref{eq:martin_eq}) cannot be solved
analytically in terms of elementary functions, but due to the singular
perturbations (terms involving $\epsilon$), it is even difficult to
solve numerically. Both of these difficulties can be conveniently
resolved with asymptotic
analysis~\cite{Newman:65,Newman:66,Bender:78}. The idea is to
construct a uniformly valid approximation for all $x$ (in the limit
$\epsilon \rightarrow 0$) by adding the leading order solution to the
``outer problem'' in the neutral bulk region $\epsilon \ll x \ll
1-\epsilon$ to the leading order solutions to the ``inner problems''
in the charged diffuse layers $0 \leq x \ll \epsilon$ and $1-\epsilon
\ll x \leq 1$ and subtracting the overlapping parts through asymptotic
matching, which is valid well below the diffusion-limited current
$|j|=1$.  As emphasized by Newman~\cite{Newman:65}, the asymptotic
approach used here reduces to the familiar Gouy-Chapman
(Poisson-Boltzmann) theory of the diffuse layer in the special case
$\delta=0$, but it is more easily generalized to the time-dependent
case. We also study for the first time the novel nonlinear effects of
the coupled Stern and Butler-Volmer boundary conditions when
$\delta>0$, which include the possibility of a reaction-limited
current.

\subsection{Leading Order Solution of the Outer Problem}

\begin{figure}
\vspace*{1.5cm}
\epsfxsize= 13.0cm
\centerline{\epsfbox{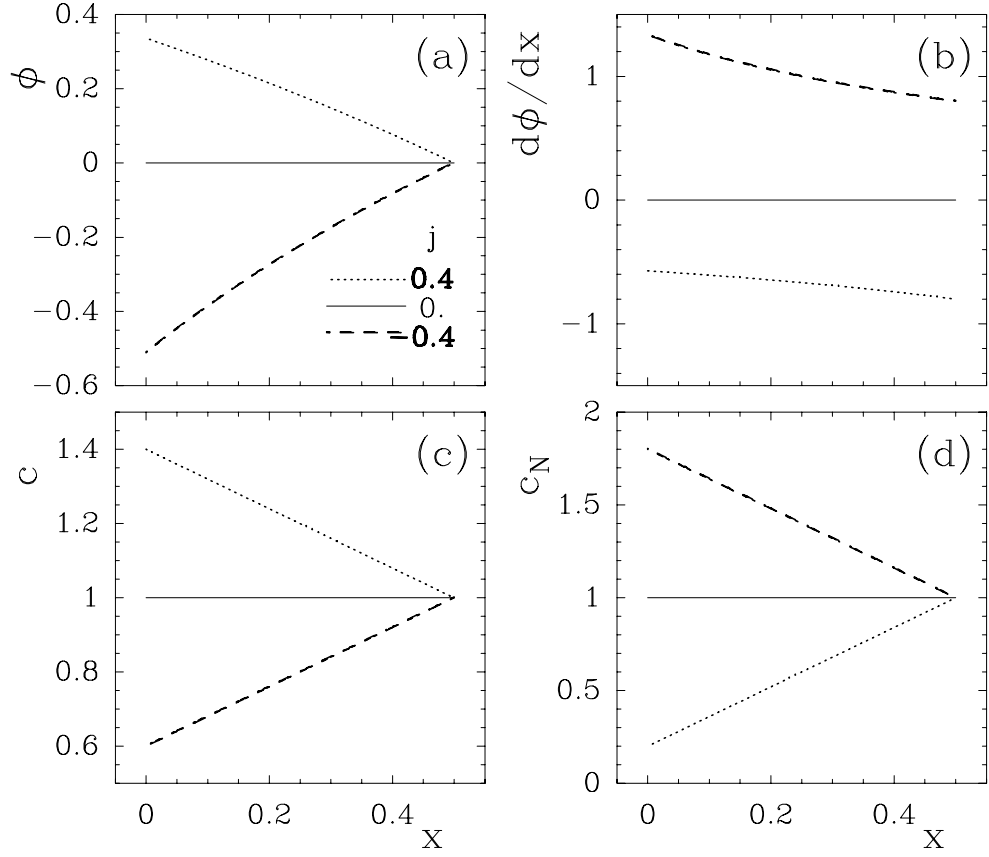}}
\vspace*{0.5cm}
\smallskip
  \caption{Steady-state profiles of $\phi$ (a), $d\phi/dy$ (b), $c$
(c), and $c_{N}$ (d) at leading order in the ``outer'' (bulk
electroneutral) region for $j=0.4, 0$ and $-0.4$ and $\gamma =
1$. $\alpha_O = \alpha_R = 1/2$  }
\label{profiles_neutral}
\end{figure}

In the outer region, we set $\epsilon=0$ in (\ref{eq:martin_eq}) to
derive the leading order approximations
\begin{subeqnarray}
\partial_x \phi & = & 2j/( 2jx - c_o ) + O(\epsilon)  \\
c(x) & = & c_o - 2jx + O(\epsilon) \\ 
\rho(x) & = & O(\epsilon^2) 
\label{full_system_neutral}
\end{subeqnarray}
from (\ref{eq:c}) and (\ref{eq:poisson}), where the integration
constant $c_o$ can be evaluated using the conservation of anions over
the cell $\int_0^1 c_A(x,t)dx=1$ at zeroth order in $\epsilon$.
\begin{equation}
c_o = 1+j
\end{equation}
since $q(t) = O(\epsilon)$. Note that the outer region is neutral
($\rho=0$) at leading order, as is commonly assumed in
electrochemistry. The bulk potential profile at leading order is
obtained by a simple integration
\begin{equation}
\phi(x) - \phi_{e} = -\dphii + \log\left[(c_o - 2jx)/c_o\right] +
O(\epsilon)
\end{equation}
where $\dphii = \phi_e - \phi(0)$ is the total (leading-order)
potential drop across the interface consisting of contributions from
the diffuse (Debye) and compact (Stern) layers
\begin{equation}
\dphii = \dphid + \dphis
\end{equation}
which must be calculated by asymptotic matching with the inner
problem. (Note again that $\dphid$ is the familiar zeta potential.)
The equation for the neutral species concentration is also easily
integrated, writing the conservation over the cell of the {\em sum} of
the cation and neutral species numbers at zeroth order in $\epsilon$.
\begin{equation}
c_{N}  = \gamma - 2j(1-2x) + O(\epsilon).
\end{equation}
Typical bulk concentration, electric field and potential profiles are
shown in Fig.~\ref{profiles_neutral}. (We set $\gamma=1$ in all the
numerical results presented here.) The potential profile $\phi$ is
slightly curved (logarithmic dependence on $x$), and the electric
field $\partial_x \phi$ is not constant since the migration term is a
nonlinear function of the concentration $c$ (which varies linearly in
the bulk when $j\ne 0$).  The profiles of $\phi$ and $\partial_x \phi$
corresponding to opposite values of $j$ (e.g. 0.4 and -0.4) are not
symmetric across $x$-axis because the product of the reduction of
cations is neutral. On the other hand, the profiles of $c$ and $c_{N}$
are symmetric when the current is reversed.

\subsection{Leading Order Solution of the Inner Problem}

The leading order approximation for the diffuse layer potential at the
$x=0$ electrode is obtained by changing variables to the inner
coordinate $y = x/\epsilon$ in (\ref{eq:martin_eq})
\begin{equation}
- \partial^3_{y} \phi + \frac{1}{2}  \left(
    \partial_y \phi \right)^3 + c_o \partial_y \phi = O(\epsilon).
\label{eq:martin_diffuse}
\end{equation}
where we ignore the $O(\epsilon)$ terms.  Multiplying by $2\ 
\partial^2_y \phi$ and integrating (using the matching conditions
$\partial_y\phi, \partial^2_y \phi \rightarrow 0$ as $y \rightarrow
\infty$) we obtain
\begin{equation}
\partial_y^2 \phi = - \partial_y \phi \sqrt{c_o + (\partial_y\phi)^2/4}
\end{equation}
which is a first-order separable equation for $\partial_y\phi$. The
next integration can be performed with an hyperbolic substitution
$\partial_y\phi = \pm 2 \sqrt{c_o} \csch(u)$ (where $\csch(u) =
1/\sinh(u)$), which yields the trivial equation $\partial_y u =
\sqrt{c_o}$, whose solution is $u = \sqrt{c_o}(y+K)$ for some constant
$K$.  Therefore, the electric field in the diffuse layer is
\begin{equation}
\partial_y \phi = \pm 2\sqrt{c_o} \csch\left(\sqrt{c_o}(y+K)\right) +
O(\epsilon)
\label{eq:Ediff}
\end{equation}
where $K$ is related to the Stern-layer voltage through the 
boundary condition at $y=0$
\begin{equation}
\dphis = \mp 2 \delta \sqrt{c_o} \csch(\sqrt{c_o} K).
\label{eq:SK}
\end{equation}
The upper sign refers to potentials below the potential of zero charge
(``p.z.c.'') with $\dphid, \dphis < 0$, and the lower sign to potentials
above p.z.c.  

Transforming to the inner coordinate in (\ref{eq:c}) and
(\ref{eq:poisson})
\begin{subeqnarray}
c &=& c_o + \frac{1}{2} \left( \partial_y \phi \right)^2 + O(\epsilon)
\\
\rho &=& -\partial^2_y \phi + O(\epsilon) 
\end{subeqnarray}
and substituting (\ref{eq:Ediff}) for the electric field, we obtain
the leading order concentration and charge density profiles
\begin{subeqnarray}
c &=& c_o\left[ 1 + 2\csch^2\left(\sqrt{c_o}(y+K)\right)\right] +
O(\epsilon)\\ 
\rho &=&  \pm 2 c_o \csch\left(\sqrt{c_o}(y+K)\right)
\coth\left(\sqrt{c_o}(y+K)\right)  +
O(\epsilon)
\end{subeqnarray}
The constant of integration $K$, or equivalently the Stern layer
voltage $\dphis$, is determined by solving a transcendental equation
provided by the Butler-Volmer kinetic boundary condition at $y=0$ (see
below)
\begin{equation}
j = -k_O ( c + \rho)\exp\left(-\alpha_O \dphis\right) + k_R (\gamma -
2j) \exp\left( \alpha_R \dphis \right)
\label{eq:K_stern}
\end{equation}
where $c(0) + \rho(0) = c_C(0)$ is a (generalized) Frumkin
correction~\cite{Newman:91}.

The expression for the electric field (\ref{eq:Ediff}) can be
integrated analytically to obtain the diffuse layer potential profile
at leading order
\begin{equation}
\phi_{e} -\phi = \Delta \phi_i + 4 \tanh^{-1}\left( e^{-\sqrt{c_o}(y+K)}
 \right)
\label{eq:GC}
\end{equation}
which is essentially the Gouy-Chapman solution to the
Poisson-Boltzmann equation, although our boundary conditions
(specifying $K$) are substantially different from the classical
theory, which corresponds to the limit $\delta\rightarrow 0$. To
clarify this connection, we let $\psi = \Delta \phi_i + \phi$, and
observe (following some algebra) that the solution (\ref{eq:GC})
satifies
\begin{equation}
\partial_y \psi = 2 \sqrt{c_o} \sinh(\psi/2)
\label{eq:GCE}
\end{equation}
and
\begin{equation}
\partial_y^2 \psi = c_o \sinh(\psi)
\label{eq:PB}
\end{equation}
which is the Poisson-Boltzmann equation. (Changing the current simply
changes the bulk concentration $c_o=1+j$.) The Smyrl-Newman equation
(\ref{eq:martin_eq}) is more general than the Poisson-Boltzmann
equation, but the former is equivalent to the latter below the
diffusion-limited current~\cite{Bazant:00}.  Combining (\ref{eq:SK})
and (\ref{eq:GCE}), we obtain a relation between the Stern and diffuse
layer voltages
\begin{equation}
\dphis = 2 \delta \sqrt{c_o} \sinh(\dphid/2)
\label{eq:ds}
\end{equation}
and from (\ref{eq:GC}) the constant $K$ can be related to the diffuse
layer voltage
\begin{equation}
\tanh(\dphid/4) = e^{-\sqrt{c_o}K}.
\end{equation}
To determine $K$ we must solve the transcendental system (\ref{eq:SK})
and (\ref{eq:K_stern}) numerically.

\subsection{Application of the Electrochemical Boundary Conditions}

\begin{figure}
\vspace*{1.5cm}
\epsfxsize= 12.0cm
\centerline{\epsfbox{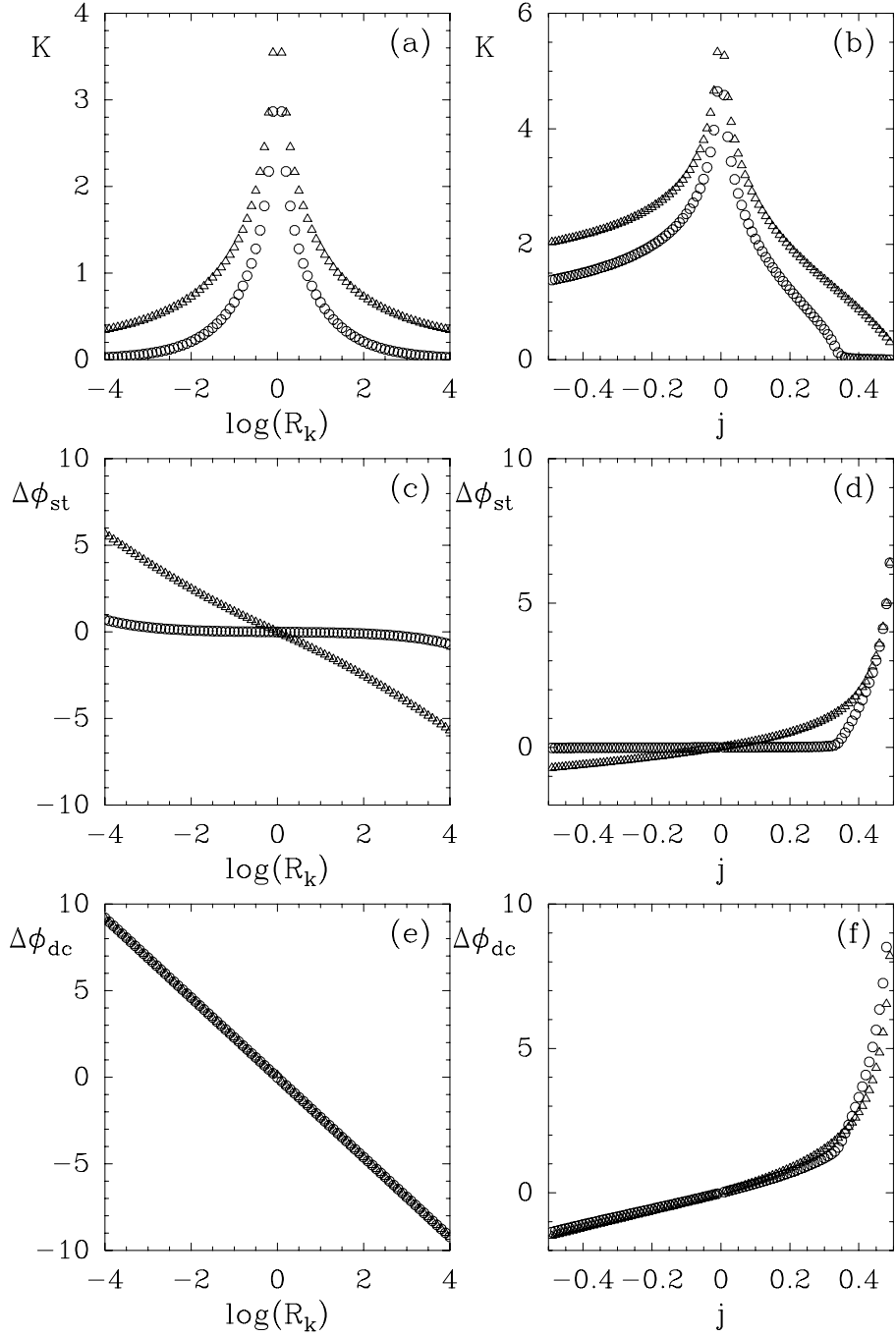}}
\vspace*{0.5cm}
\smallskip
\caption{(a) $K$ versus $R_k= k_r/k_o$ in lin\--log representation, 
($j=0$). (b) $K$ versus $j$ for $R_k=1$. (c) $\Delta \phi_{s}$
versus $R_k$ in lin\--log representation, 
($j=0$).  (d) $\Delta \phi_{S}$ versus $j$ for $R_k=1$. 
(c) $\Delta \phi_{i}$
versus $R_k$ in lin\--log representation, 
($j=0$), $\gamma =1 $.  
(d) $\Delta \phi_{i}$ versus $j$ for $R_k=1$. 
Two different ratios $\delta$
are illustrated in
the six panels: circles $\delta =0.01$, triangles
$\delta =1 $.  $\alpha_O = \alpha_R = 1/2$ 
}
\label{new_k_fi_dc_vs_kr_ko_j}
\end{figure}

\begin{figure}
\vspace*{1.5cm}
\epsfxsize= 12.0cm
\centerline{\epsfbox{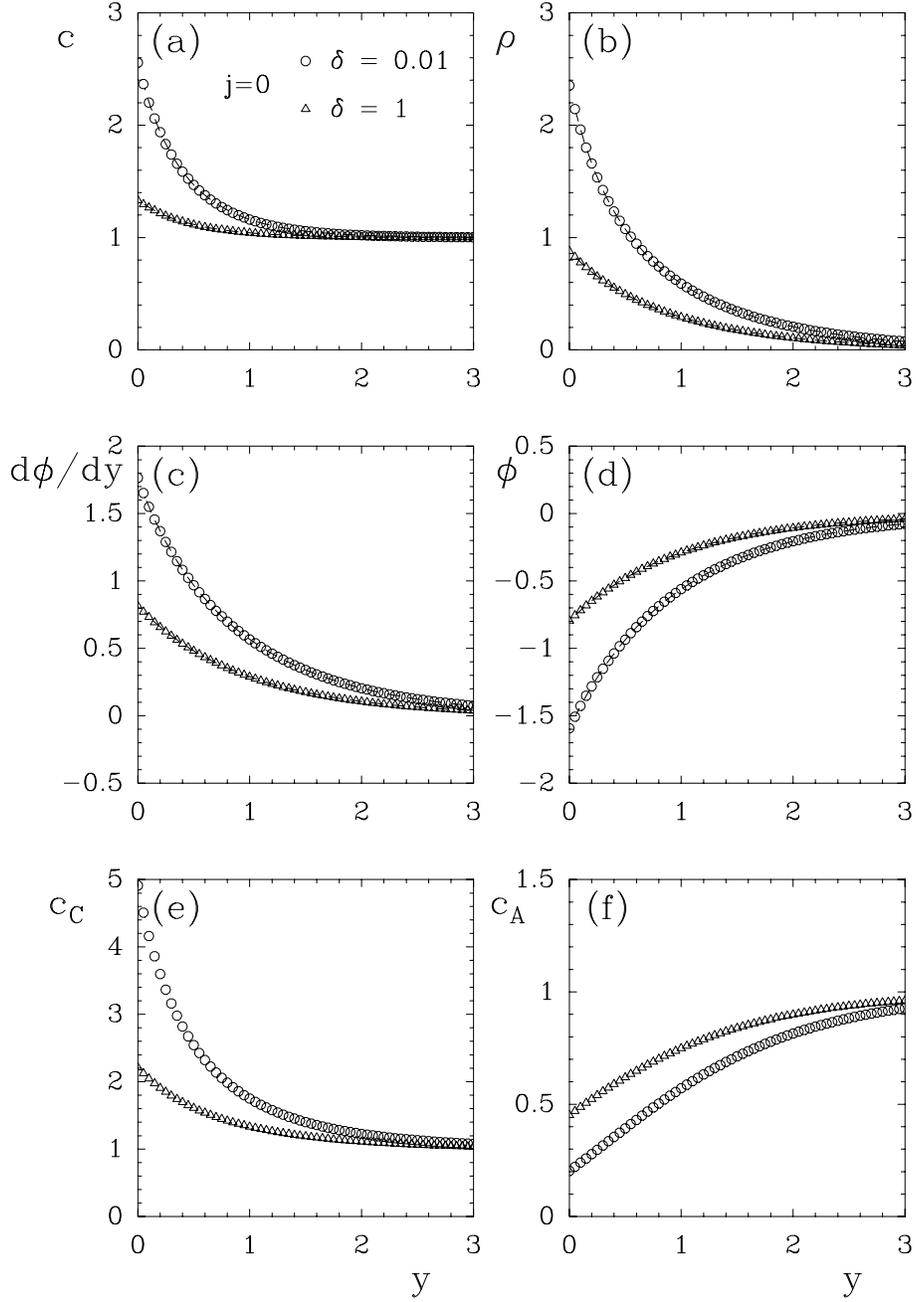}}
\vspace*{0.5cm}
\smallskip
\caption{Steady-state profiles of $c$ (a), $\rho$ (b), $\partial_y
\phi$ (c), $\phi$ (d), $c_C$ (e) and $c_A$ (f) at leading order in the
``inner'' (diffuse) layer at $y=0$ for $j=0$, $R_k = 5$, ($k_r =5 $
and $k_o=1$), $K=0.972$ for $\delta = 0.01$ (circles) and
$K=1.629$ for $\delta=1$ (triangles), For comparison the
Poisson\--Boltzmann's profiles are reported as dashed lines 
which correspond to the limit $\delta=0$.
}
\label{new_conc_rho_dfi_vs_y_a}
\end{figure}

\begin{figure}
\vspace*{1.5cm}
\epsfxsize= 12.0cm
\centerline{\epsfbox{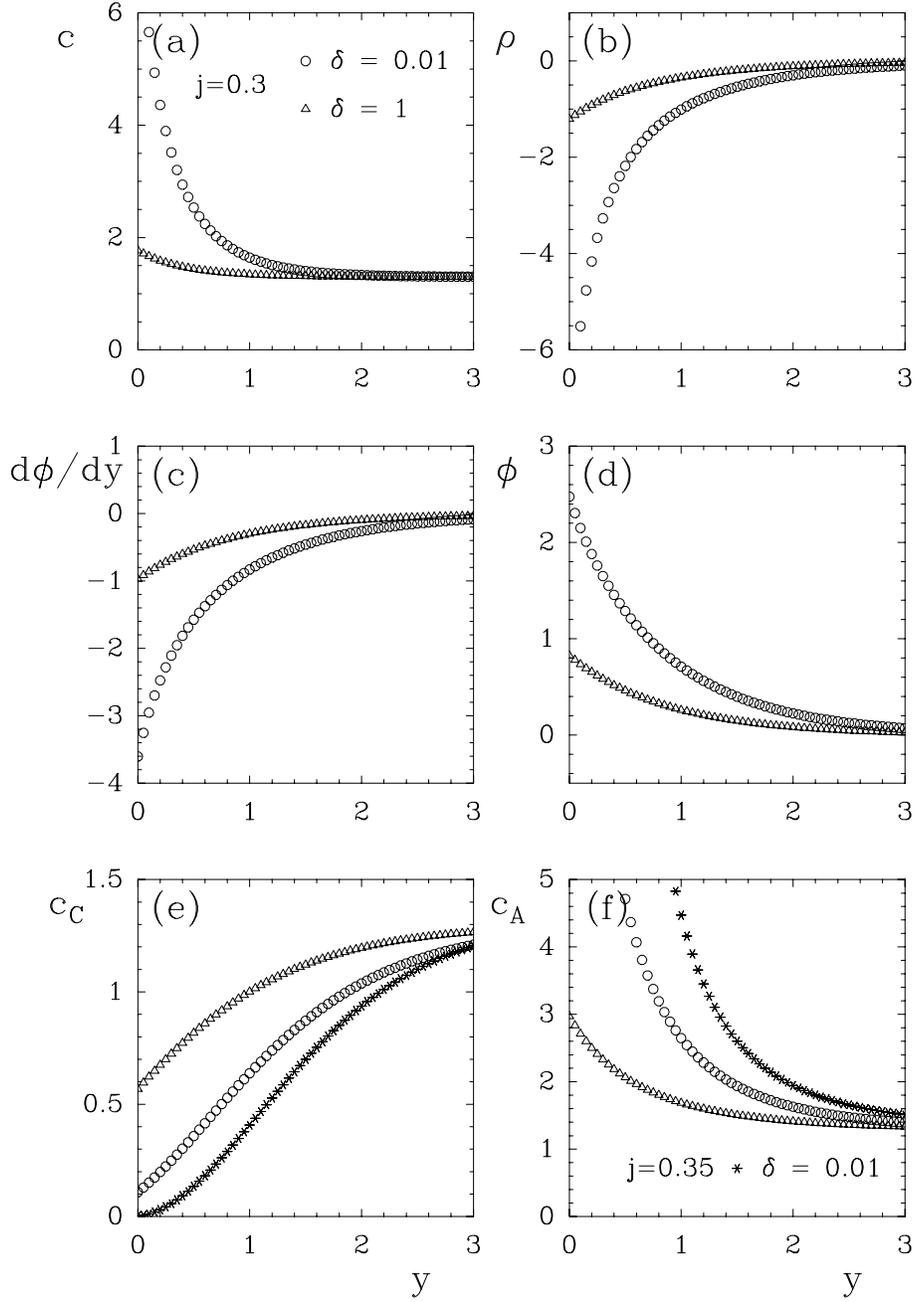}}
\vspace*{0.5cm}
\smallskip
\caption{Steady-state profiles of $c$ (a), $\rho$ (b), $\partial_y
\phi$ (c), $\phi$ (d), $c_C$ (e) and $c_A$ (f) in the diffuse layer at
leading order for $j=0.3$, $R_k=1$, $K=0.523$ for $\delta = 0.01$
(circles) and $K=1.394$ for $\delta=1$ (triangles). }
\label{new_conc_rho_dfi_vs_y_b}
\end{figure}

The behavior of $K$ as a function of the ratio of the kinetics
constants $R_k = k_R/k_O$ and the current density $j$ is reported in
Fig.~\ref{new_k_fi_dc_vs_kr_ko_j} (with $k_R=1$ held constant).  From
(\ref{eq:SK}) it is clear that large values of $K$ correspond to small
values of $\dphis$ when $j=0$, i.e. the symmetric case $c_o=1$.  The
introduction of a nonzero value of $j$ breaks the symmetry of the $K$
versus $\log(R_k)$ curve.  The plot of $K$ versus $j$ in
Fig.~\ref{new_k_fi_dc_vs_kr_ko_j}(b) illustrates the dissymmetry of
the redox couple, since when the current is positive the oxidation of
neutral species produces positively charged cations which accumulate
in the diffuse layer, thus hindering the negative charging of the
diffuse-layer. In Fig.~\ref{new_k_fi_dc_vs_kr_ko_j} we also display
the Stern and total interfacial voltages, $\dphis$ and $\dphii =
\dphis + \dphid$.  We observe that when $j$ increases beyond 0.3,
$\dphis$ and $\dphii$ reach arbitrarily large values to compensate
this excess of cations (see below).  For $j=0$, the plots of $\dphis$
(Fig.~\ref{new_k_fi_dc_vs_kr_ko_j}(c)) and $\dphii$
(Fig.~\ref{new_k_fi_dc_vs_kr_ko_j}(e)) are symmetric with respect to
$R_k=1$.  This behavior is confirmed by the plot of the different
spatial profiles $c$, $\rho$, $\partial_y \phi$ and $\phi$, for two
different values of $j$, in Figs.~\ref{new_conc_rho_dfi_vs_y_a} and
\ref{new_conc_rho_dfi_vs_y_b}.

In Fig.~\ref{new_conc_rho_dfi_vs_y_a} we focus on the case of zero
current. When the kinetic constants are equal $R_k=1$, this correspond
to the p.z.c., but when $R_k \ne 1$, the system builds a charged
diffuse layer to equilibrate the oxidation and reduction currents
on each electrode.  For comparison, we have also added in (dashed
lines) the stationary Poisson-Boltzmann solution for $j=0$, which
corresponds to choosing $K$ such that $\delta=0$ in (\ref{eq:GC}).

In Fig.~\ref{new_conc_rho_dfi_vs_y_b}, a positive current $j=0.3$ has
been chosen, with symmetric kinetic constants ($R_k = 1$).  In this
case, the system builds a charged diffuse layer to favor either
reduction (negative Faradaic current) or oxidation (positive Faradaic
current). In each figure, two values of $\lambda_D$ have been chosen,
one corresponding to a weak electrolyte concentration $\lambda_D = 100
\lambda_{S}$, where the diffuse layer has a strong effect on
the charge distribution and potential profiles, and another
corresponding to a strong electrolyte concentration 
with $\lambda_D=\lambda_S$,
where there is a much more significant potential drop in the Stern
layer (and hence less influence from the diffuse layer).  We note
that for $j=0.3$ the charge and electric field profiles shown in
Fig.~\ref{new_conc_rho_dfi_vs_y_b} start to diverge at $y=0$,
particularly so at the small value of $\delta=1/100$. 

\begin{figure}
\vspace*{1.5cm}
\epsfxsize= 7.cm
\centerline{\epsfbox{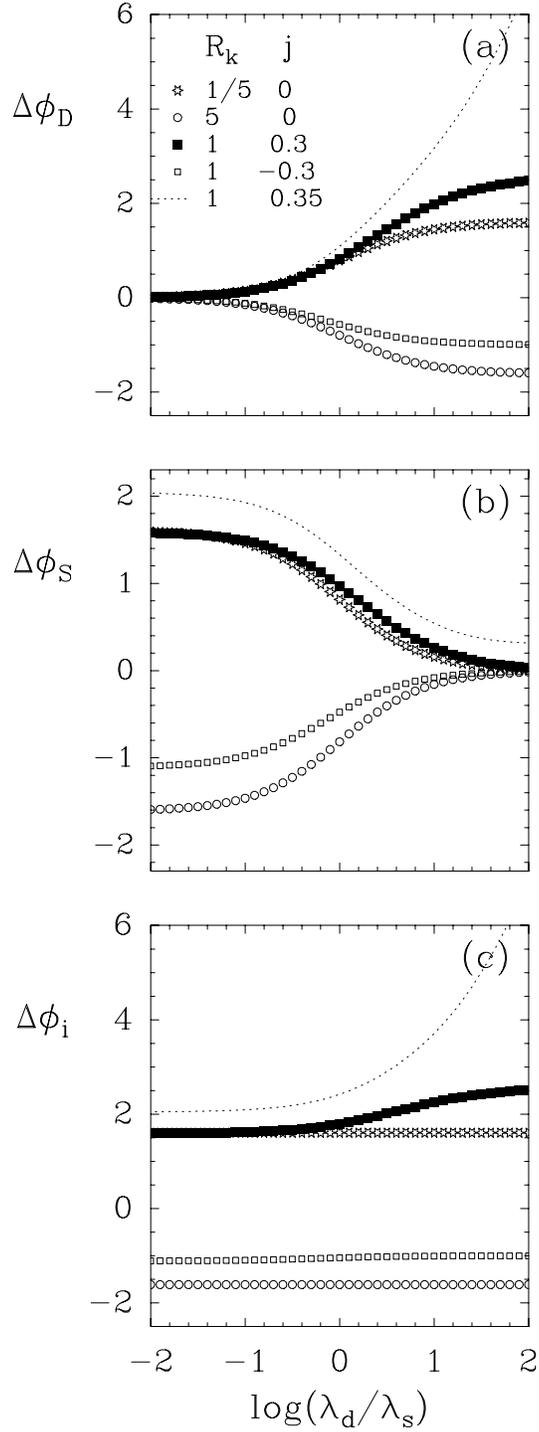}}
\smallskip
\caption{Dependence of the potential differences $\Delta
  \phi_{D}$ (diffuse layer), $\Delta
  \phi_{S}$ (Stern's layer), $\Delta
  \phi_{i} = \Delta \phi_{D} + \Delta \phi_{S}$ (double layer) with
  $-\log \delta$. The different parameters $R_k$ and
  $j$ corresponding to each plot are inserted in panel (a).
}
\label{new_fid_fist_vs_ld}
\end{figure}

To better understand this unexpected behavior observed for $j>0.3$, we
plot the influence of the ratio $\lambda_{D} /\lambda_{S} =
\delta^{-1}$ on the potential drops of the diffuse layer $\dphid$,
Stern's layer $\dphis$ and the whole double layer $\dphii = \dphid +
\dphis$ in Fig.~\ref{new_fid_fist_vs_ld}. We notice that an exchange
of potential drop between Stern's layer and the diffuse layer occurs
when the size of the diffuse layer becomes larger than
$\lambda_S$. The potential drops in Stern's layer and diffuse layer
compensate exactly when $j=0$, whereas the potential drop in the
diffuse layer diverges when $\lambda_{D}$ increases at a nonzero
current larger than $0.3$.

\subsection{Theoretical Polarograms}

\begin{figure}
\vspace*{1.5cm}
\epsfxsize= 9.5cm
\centerline{\epsfbox{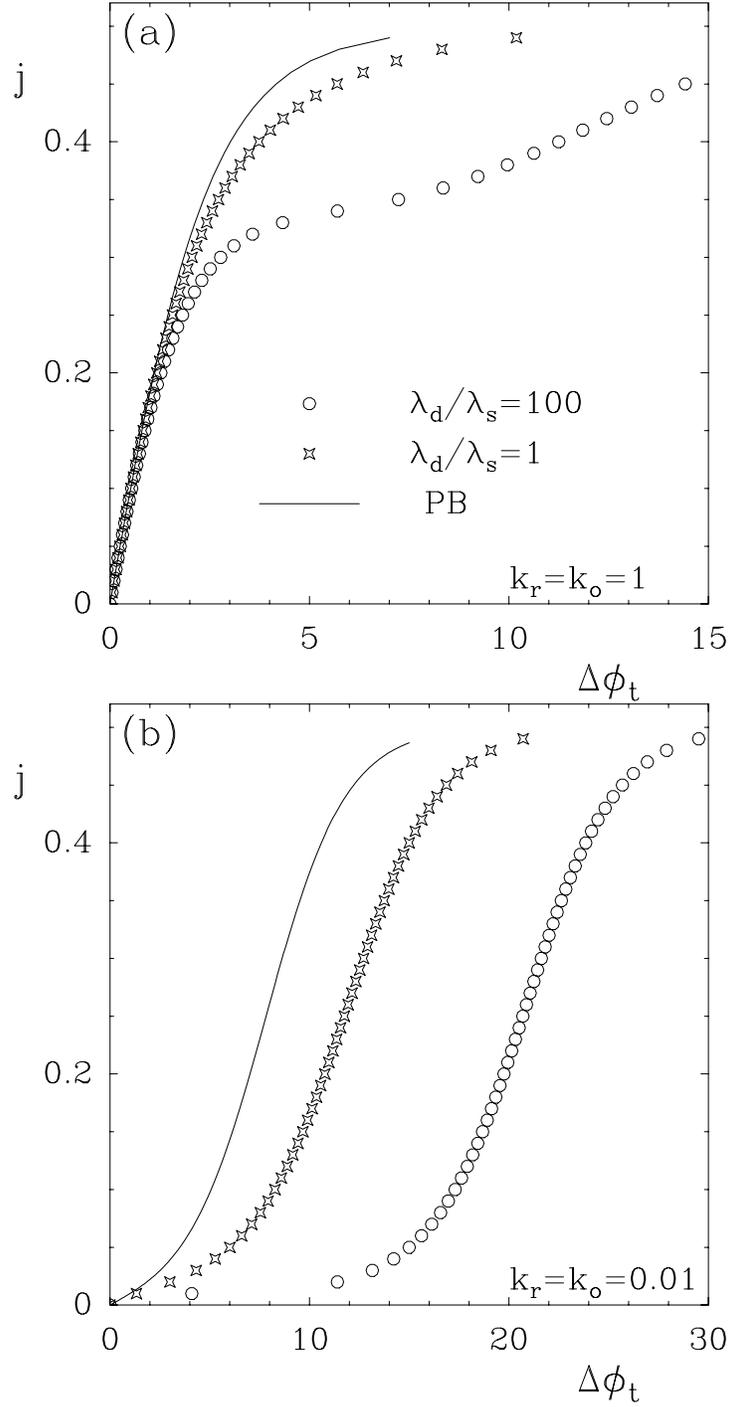}}
\smallskip
\caption{Polarograms obtained from the asymptotic analysis
for   (a) $k_R=k_O = 1$ and (b) $k_R=k_O = 0.01$. }
\label{new_j_vs_dfi_total}
\end{figure}

For the electrochemist, polarogramms bear a great significance. In the
present study of stationary regimes in finite size cells such a
representation of current $j$ versus the total voltage between the
reference and working electrodes $\Delta \phi_{tot} = \phi_{e} -
\phi_{ref}$ is shown in Fig.~\ref{new_j_vs_dfi_total}. This includes
leading order voltage drops calculated for both double layers and the
bulk electroneutral region.  It confirms again than for $j$ positive
large than 0.3, a limitation of the current occurs which is very
different from diffusion limited current which are usually encountered
in electrochemical cells. The standard current\--potential curve
(``PB'', plotted as a dotted line in Fig.~\ref{new_j_vs_dfi_total})
shows a limiting current equal to $\gamma/2$ which occurs when $c_{N}
\rightarrow 0$ at the working electrode surface.  When $j>0$ the
production of cations increases due to oxidation of neutral species,
to maintain the imposed current, the system usually increases $\Delta
\phi_{S}$ to compensate this excess.  When $\delta =
\lambda_{S}/\lambda_{D}$ becomes much smaller than 1, the potential
drop in the diffuse layer increases much more rapidly than the
potential drop in the Stern layer, there is an enhancement of the
electric field in the diffuse layer which tends to strip the cations
from the anode. For a given current density ($j>0.3$ in this case) the
interfacial cation concentration tends to zero. This limitation of the
current, however, is not due to transport but rather the reaction
kinetics coupled to the diffuse layer charging.

The significance of $j=0.3$ in our numerical results is most easily
understood in the limit of a very weak electrolyte with $\lambda_D \gg
\lambda_S$ ($\delta \rightarrow 0$) where it corresponds to a
``reaction-limited current'', generally given by
\begin{equation}
j_{RL} =  \frac{k_R \gamma}{1 + 2 k_R}
\label{eq:jRL}
\end{equation}
whenever this value is smaller than the Nernst diffusion-limited
current $\gamma/2$, when $c_N(0)=0$.  (Recall that we have chosen $k_R
=\gamma = 1$, so that $j_{RL} = 1/3$ and $\gamma/2=1/2$.)  Instead of
being caused simply by transport limitations, the reaction-limited
current $j_{RL}$ is determined by the electrochemical boundary
conditions. Note that when $\delta=0$ (no Stern layer), and the
Butler-Volmer equation (\ref{eq:K_stern}) takes the simpler form
\begin{equation}
j = - k_O(c+\rho) + k_R(\gamma-2j),
\end{equation}
and in this case, a limitation of positive current occurs because the
first term representing the cathodic current is negative (but small)
while the positive second term representing the anodic current
decreases with applied current as the concentration of neutral species
is reduced (but not necessarily depleted) at the electode. When
$\delta>0$ (however small), the situation is quite different because
the system can avoid the reaction limitation in (\ref{eq:K_stern}) by
producing a large positive Stern-layer voltage $\dphis$
\begin{equation}
\dphis \sim \frac{1}{\alpha_R} \log\left(\frac{j}{k_R(\gamma-2j)}\right)
\end{equation}
which suppresses the negative cathodic current and exponentially
enhances the positive anodic current enough to maintain the applied
total current. This also produces a large diffuse layer voltage for $j
> j_{RL}$ according to (\ref{eq:ds}), although this effect is reduced
as $\delta$ is increased.

The actual limiting current in the system is controlled by the smaller
of the reaction-limited current (\ref{eq:jRL}) and the diffusion
limited current $|j|=\gamma/2$. Note that for negative currents one
observes a diffusion limiting current at $j=-\gamma/2$ which
corresponds to the annulation of the neutral species concentration at
the counter electrode. Therefore, the system never reaches the
classical limiting value $j=-1$ which would be obtained with a
strongly supporting electrolyte (purely diffusing system).

\section{Analysis of the Linear Response to an AC Current Modulation}
\label{sec:imp}

In this section we derive the leading-order response of our simple
electrochemical system to small amplitude AC current perturbations
(around a steady nonzero DC current) using the same mathematical
methods as in previous section, now applied to the Laplace transforms
of the linearized equations.  There have been very few attempts in the
existing literature~\cite{MacDonald:74,MacDonald:79,MacDonald:87} to
compute the impedance spectra for electrochemical systems with weak
electrolytes (and large Debye lengths) in the absence of a supporting
electrolyte. The modeling of voltammetric curves has received more
attention~\cite{Bamford:86,Oldham:92,Bento:98,Bento:98b,Amatore:99},
although most of these studies either neglect the diffuse layer
entirely or assume its ionic concentration profiles are those of
equilibrium. In this section we show that the diffuse-layer cannot
always be approximated by an additional fixed capacitance in parallel
with the Stern-layer capacitance because its dynamics can strongly
influence the net interfacial impedance at intermediate frequencies.

To compute the response of the electrochemical cell to small applied
AC current modulation $\delta j$ of a DC current $\overline{j}$
\begin{equation}
j  =  \overline{j} + \delta j
\end{equation}
we add a corresponding small modulation to each of the variables
\begin{subeqnarray}
c & = & \overline{c} + \delta c \\ 
c_{N} & = & \overline{c}_{N} + \delta c_{N} \\
\rho & = & \overline{\rho} + \delta \rho \\
\phi & = & \overline{\phi} + \delta \phi \\
\label{modulation}
\end{subeqnarray}
about the leading-order approximation of the steady-state solution
described above (denoted by an overline), and linearize the
time-dependent model of equations (\ref{eq:eqns}) around the stationary
state to obtain the following system:
\begin{subeqnarray}
\partial_{\tau} \delta c & = & \partial^2_x \delta c + 
  \partial_x  \left( \overline{\rho} \partial_x
  \delta \phi + \delta \rho \partial_x \overline{\phi} \right) \\
\partial_{\tau} \delta \rho & = & \partial^2_x \delta \rho + 
  \partial_x  \left( \delta c \partial_x
  \overline{\phi} + \overline{c} \partial_x \delta \phi \right) \\
\partial_{\tau} \delta c_N & = & \partial^2_x \delta c_N \\
\delta \rho & = & -\epsilon^2 \partial^2_x \delta \phi  
\label{transport_modulation}
\end{subeqnarray}
The electrochemical boundary conditions at both electrodes ($x=0,1$)
are also linearized about the steady-state
\begin{subeqnarray}
-2 \delta j_F & = & \partial_{x} \delta c + 
  \overline{\rho} \partial_x
  \delta \phi + \delta \rho \partial_x \overline{\phi}  \\
-2 \delta j_F & = & \partial_{x} \delta \rho + 
  \delta c \partial_x
  \overline{\phi} + \overline{c} \partial_x \delta \phi \\
4 \delta j_F & = & \partial_{x} \delta c_N \\
\delta j_F & = & \left[k_O \alpha_O (\overline{c} + \overline{\rho})
 \exp(-\alpha_O \overline{\dphis}) 
 + k_R \alpha_R c_{N} \exp(\alpha_R \overline{\dphis})\right] \delta\dphis \\
& & - k_O (\delta c + \delta\rho) \exp(-\alpha_O \overline{\dphis}) 
+ k_R \delta c_N \exp(\alpha_R \overline{\dphis}) \\
\delta\dphis &=& \left\{ \begin{array}{ll} 
- (\delta\cdot\epsilon) \partial_x \delta \phi & \ \mbox{ at } x=0 \\
(\delta\cdot\epsilon) \partial_x \delta\phi & \ \mbox{ at } x=1
\end{array} \right.
\label{eq:modbc}
\end{subeqnarray}
where the Stern-layer voltage is decomposed as the other variables
$\dphis = \overline{\dphis} + \delta\dphis$. In the time-dependent
setting, the current modulation is broken into separate contributions
from the Faradaic conduction and displacement current densities
\begin{equation}
\delta j = \delta j_F + \delta j_d
\end{equation}
where
\begin{subeqnarray}
\delta j_F & = & -\frac{1}{2}\left(\partial_x \delta\rho + \delta c
\partial_x \overline{\phi} + \overline{c}\partial_x \delta\phi \right) \\
\delta j_d &=& - \frac{1}{2}\epsilon^2 \partial_\tau \partial_x
\delta \phi 
\end{subeqnarray}
The unperturbed DC current is entirely Faradaic ($\overline{j} =
\overline{j_F}$, $\overline{j_d}=0$), but the displacement current becomes
important in the AC modulation, especially at large frequencies and
small length scales.

To compute the impedance of the electrochemical cell, we consider a 
small amplitude AC perturbations of the form 
\begin{equation}
\delta j = j_o \sin (\omega t)
\end{equation}
and take the Laplace transform in time of the system of transport
equations (\ref{transport_modulation})
\begin{subeqnarray}
s \tilde{\delta c} & = & \partial^2_x \tilde{\delta c} + 
  \partial_x  \left( \overline{\rho} \partial_x
  \tilde{\delta \phi} + \tilde{\delta \rho} \partial_x \overline{\phi}
  \right) \\ 
s \tilde{\delta \rho} & = & \partial^2_x \tilde{\delta \rho} + 
  \partial_x  \left( \tilde{\delta c} \partial_x
  \overline{\phi} + \overline{c} \partial_x \tilde{\delta \phi} \right) \\
s \tilde{\delta c}_n & = & \partial^2_x \tilde{\delta c}_n \\
\tilde{\delta \rho} & = & -\epsilon^2 \partial^2_x \tilde{\delta \phi} \\ 
\label{transport_modulation_laplace}
\end{subeqnarray}
where $s = i\omega L^2/D$ is the Laplace variable conjugate to time
(the dimensionless forcing frequency, or imaginary decay rate) and
tilde accents are used denote the transformed variables $\tilde{\delta
f} = {\mathcal L} \delta f$. Since they do not depend explicitly on
time, the interfacial boundary conditions (\ref{eq:modbc}) are the
same for the transformed variables as for the original ones.  Note
that the displacement current modulation takes the form
\begin{equation}
\tilde{\delta j_d} = - \frac{1}{2} \epsilon^2 s \partial_{x}
  \tilde{\delta \phi}
\label{eq:jds}
\end{equation}
in Laplace space. Once we solve these equations for the perturbed
variables, the (complex) impedance of the cell is calculated as the
ratio of the voltage response versus the current modulation in Laplace
space
\begin{equation}
Z(\omega) = \frac{\tilde{\delta \Delta \phi_{tot}}}{\tilde{\delta j}}
\end{equation}
where $\tilde{\delta \Delta \phi_{tot}}$ is the total voltage
modulation (in Laplace space) between the working electrode and the
reference electrode.

The analysis begins by combining the transport equations
(\ref{transport_modulation_laplace}) following the same steps as in the
derivation of the Smyrl-Newman equation above, which leads to a
system  of two linear equations for $\tilde{\delta \phi}$ and
$\tilde{\delta c}$: 
\begin{subeqnarray}
-s \epsilon^2 \partial^2_x \tilde{\delta \phi} & = & -\epsilon^2
\partial^4_x \tilde{\delta \phi} + \partial_x \left( 
\tilde{\delta c} \partial_x \overline{\phi} + \overline{c} \partial_x
\tilde{\delta \phi} \right) \\
s \tilde{\delta c}  & = & \partial^2_x \tilde{\delta c} - \epsilon^2 
\partial^2_x \left( \partial_x \overline{\phi} \partial_x
  \tilde{\delta \phi}\right) 
\label{Supereqphix}
\end{subeqnarray}
Although these equations are linear, they cannot be solved analytically in
terms of elementary functions, and they also possess singular
perturbations which hinder numerical progress. As in the steady-state
case, asymptotic analysis in the limit $\epsilon \rightarrow 0$ once
again allows us to solve the leading-order problem by decomposing the
cell into a bulk electroneutral region which must be asymptotically
matched with charged diffuse boundary layers. The resulting leading
order equations for the inner and outer problems are difficult to
solve analytically in this case, but they present no trouble for a
numerical solution because the singular perturbations have been
removed analytically. The details of these calculations will be
presented elsewhere, but here we present a simple example of the rich
time-dependent electrochemical behavior that can be predicted by the
mathematical model.

In the framework of the asymptotic analysis, the total impedance of
the cell is conveniently broken into separate leading-order
contributions from the various regions
\begin{equation}
Z(\omega) = \frac{\tilde{\delta \dphis} + \tilde{\delta \dphid}
  + \tilde{\delta \Delta \phi_{b}}}{\tilde{\delta j_F} + \tilde{\delta j_d}}
\end{equation}
where $\tilde{\delta \dphis}$, $\tilde{\delta \dphid}$ and
$\tilde{\delta \Delta \phi_{b}}$ are the voltage modulations in
the Stern layer, the diffuse layer and the bulk
electroneutral zone, respectively.
The asymptotic analysis for the limit $\epsilon \rightarrow 0$ assumes
that the diffuse layers are stationary (quasi-steady) at leading
order, which is valid when the forcing frequency is below the Debye
frequency 
\begin{equation}
\omega \ll \frac{D}{\lambda_D^2}
\end{equation}
or in terms of the dimensionless frequency
\begin{equation}
|s| = \frac{\omega L^2}{D} \ll \frac{1}{\epsilon^2}.
\end{equation}
This does not limit the applicability of our analysis because
impedance spectroscopy experiments are conducted well below the Debye
frequency. 

It is interesting to consider the displacement current density over
this range of time scales. From (\ref{eq:jds}) we see that when $|s|
\ll 1/\epsilon^2$ or $\epsilon^2 s \ll 1$ the displacement current
$\tilde{\delta j_d}$
is always negligible in the outer electroneutral region. However, the
situation is very different in the diffuse layers, where the natural
length scale $\lambda_D$ is much smaller. In the diffuse layer, we
write the displacement current in terms of the inner coordinate
$y=x/\epsilon$, 
\begin{equation}
\tilde{\delta j_d} = -\frac{1}{2}s\epsilon \partial_y \tilde{\delta \phi}
\end{equation}
and observe that it becomes comparable to the Faradaic current when
$|s| \lambda_D \tilde{\delta \Delta \phi_{S}}/\lambda_S =
O(\epsilon^{-1})$. Since the kinetic constants ($k_O$, $k_R$) are
included in the value of $ \tilde{\delta \Delta \phi_{S}}$, they will
have strong influence on the current displacement and on the
``crossover'' frequency where the displacement and Faradaic currents
become comparable.

Note that for the case of non\--reacting system (zero kinetic
constants, and zero Faradaic current) the characteristic frequency of
Debye layer charging has recently been derived in the completely
different context of electrohydrodynamic response to a spatially
periodic applied AC voltage~\cite{Ajdari:00}:
\begin{equation}
\omega_{c} = \frac{D}{\lambda_D L} \; \; \;  .
\end{equation}
This prediction is recovered by our model in the limit where
$\lambda_{D} \gg \lambda_{S}$.

\begin{figure}
\vspace*{1.5cm}
\epsfxsize= 12.0cm
\centerline{\epsfbox{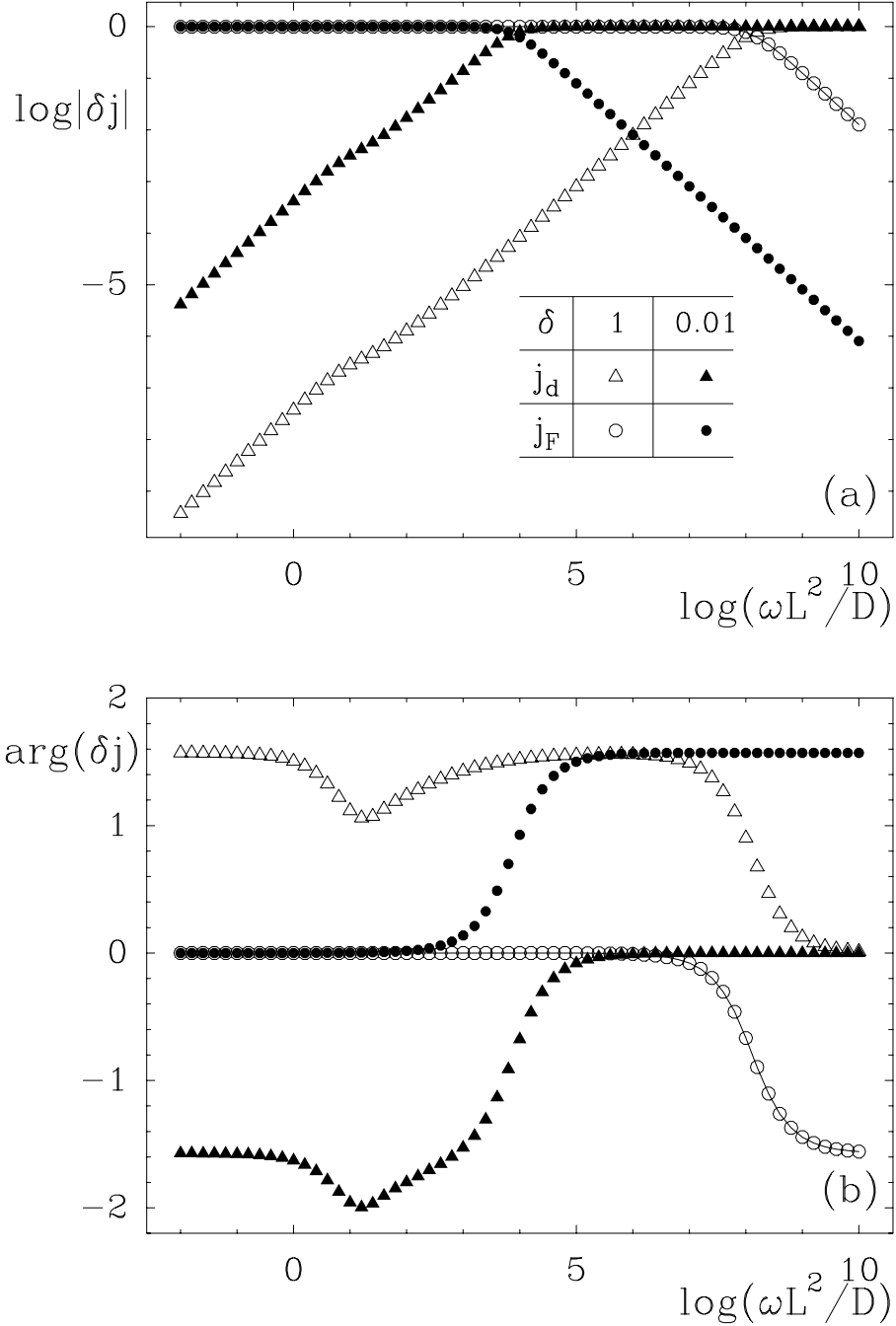}}
\vspace*{0.5cm}
\smallskip
\caption{The displacement and Faradaic current modulations versus the
dimensionless frequency $\omega L^2/D$. Results for two different
ratios $\delta=\lambda_S/\lambda_D$ are shown. (a) $\log$ of the
modulus of current modulation $\tilde{\delta j}$.  (b) argument of the
current modulation $\tilde{\delta j}$.  $\lambda_{S}=1 \AA$ and
$L=1$~cm, so $\lambda_{D}=10$nm corresponds to $\delta = 0.01$ and
$\epsilon=10^{-6}$, while $\lambda_{D}=1 \AA$ corresponds to $\delta =
1$ and $\epsilon=10^{-8}$. Other parameters are $D =
10^{-5}$cm$^{2}$s$^{-1}$, $j=0.3$, $k_R=k_O=1$, $\gamma = 1$,
$\alpha_O=\alpha_R=1/2$.  }
\label{cour_frequence}

\end{figure}

In Fig~\ref{cour_frequence}, a plot of the two current modulations
versus the AC frequency $\omega$ gives a clear insight into the
interchange of the reaction and diffuse layer charging processes with
frequency. (The amplitude of the stationary current in this example is
$j=0.3$, near the steady-state reaction-limited current $j_{RL}$). The
crossover frequency where the displacement and Faradaic current
amplitudes are comparable is strongly affected by the parameter
$\delta = \lambda_S / \lambda_D$. The larger is $\delta$, the smaller
is the diffuse layer size and the higher the frequency at which the
displacement currents dominate, given (very roughly) by
$\epsilon^{-1}$.

\begin{figure}
\vspace*{1.5cm}
\epsfxsize= 12.0cm
\centerline{\epsfbox{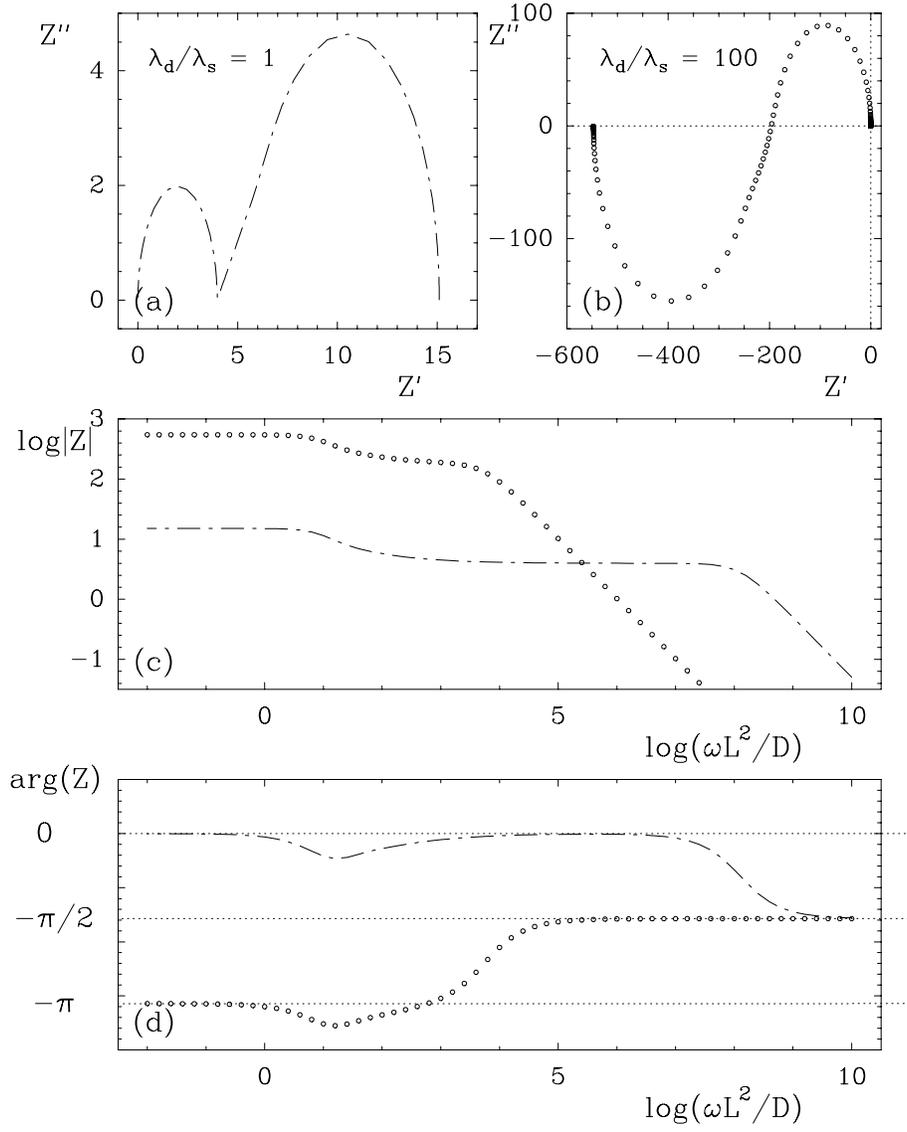}}
\vspace*{0.5cm}
\smallskip
\caption{Impedance spectra obtained from a linear stability analysis
  around the steady state $j=0.3$. (a) Nyquist plot, (b)  and (c) Bode plots
  for the modulus and argument of the total impedance $Z(s=i\omega) = 
\tilde{\Delta \phi_{tot}(s)} /\tilde{\delta j(s)}$. 
Same parameters as in Fig.~\ref{cour_frequence}}
\label{fig_impedances}
\end{figure}

The arguments of the current modulations are such that when
$\lambda_{D}$ increases the phases become shifted by $\pi$, showing an
out of phase AC voltage response. This phenomena is often termed
``negative impedance''
\cite{Koper:93b,Koper:96,Strasser:99,Krischer:99} and has already been
reported in the electrochemical literature as the source of
instabilities (stationary or oscillatory). The impedance spectra
corresponding to the current modulations reported in
Fig.~\ref{cour_frequence} are shown in Fig.~\ref{fig_impedances}. The
two examples $\delta = 1$, $\epsilon=10^{-8}$ and $\delta =0.01$,
$\epsilon=10^{-6}$ are consistent with the phase shift that we have
described in the current modulation curves. The amplitude of the
impedance also increases very dramatically when the current density
goes beyond $0.33$. As already discussed in the stationary case, when
the system reaches this ``reaction-limited current'' situation, a
strong potential drop is established in the diffuse layer, which can
explain the amplification of the electrochemical cell impedance.  We
hope to elaborate on this aspect of the theoretical predictions in
comparison with real experiments in a future communication.

\section{Conclusion}

We have performed a leading-order asymptotic analysis of a realistic
set of transport equations and boundary conditions for a three-species
electrochemical cell in which cations are reduced into a neutral
molecules in the presence of anions, in both stationary and
time-dependent nonequilibrium situations. By solving the full
nonlinear problem at leading order (numerically where necessary), we
have made the first systematic exploration of the influence of the
parameter $\delta = \lambda_S/\lambda_D$ (introduced in
Ref.~\cite{Bazant:00} to describe the coupling of the Stern and
diffuse layers) on stationary and time-dependent interfacial
kinetics. The mathematical model is able to predict very subtle
effects related to the coupling between diffuse charge and interfacial
reactions, such as the existence of both reaction-limited and
diffusion-limited currents and negative impedances with nonnegligible
displacement currents are large frequencies. The strength and novelty
of these theoretical predictions is their firm analytical basis
starting from realistic transport equations and boundary conditions.

An interesting extension of this study would be to electrochemical
reactions involving the reduction of anions (or the oxidation of
cations).  Such systems have already been the subject of many
investigations, both theoretical and experimental, but most these
studies have not considered simultanously the dynamics of the
diffusion and diffuse layers and their possible interplay during the
interfacial reactions, as we have in this work. Such situations would
complicate the present analysis signficantly because there would be
more than two charged species, and it appears that any analytical
solution of the inner problem at leading order would not be possible
in such cases. However, as long as $\epsilon$ is small ($\lambda_D \ll
L$), the mathematical machinery of asymptotic analysis can be used to
derive simplified, well-behaved equations at leading order, which
would be straightforward to at least integrate numerically.

Even restricting ourselves to the simple electrochemical system
studied here, there are number of interesting directions for future
work. Higher order corrections in $\epsilon$ could be derived (with
some numerical computation required). It would also be interesting to
solve the full system of equations with $\epsilon=O(1)$, where the
asymptotic analysis breaks down. This limit, which corresponds to very
small cells on the order of the Debye length, has increasing
importance as microelectrochemical systems reach smaller and smaller
length scales. It would also be interesting to solve the AC current
problem more globally in the nonlinear regime, where our local linear
stability analysis is no longer valid.  In order to study nonlinear
responses, such as those encountered in voltammetry or
chronoamperometric methods, the full nonlinear system of equations
would have to be solved numerically. Nevertheless, if $\epsilon$ is
small (as is usually the case) asymptotic analysis provides the
appropriate starting point for the mathematical modeling of
diffuse-charge effects in electrochemical systems.

\acknowledgements

We are very grateful to Armand Ajdari, Fran\c{c}ois Nadal and Jacek
Lipkowski for stimulating discussions.  This work has been supported
by the Centre National des Etudes Spatiales (CNES) under grants
97/CNES/071/6850 and 793/98/CNES/7315.

\bibliography{double_layer,impedances}

\begin{thebibliography}{10}

\bibitem{Frumkin:55}
A. Frumkin, Z. Electrochem {\bf 59},  807  (1955).

\bibitem{Grahame:47}
P. Delahay, Chem. Rev. {\bf 41},  441  (1947).

\bibitem{Parsons:54}
R. Parsons,  in {\em Modern Aspects of electrochemistry}, edited by J.-O.
  Bockris and B. Conway (Butteworths, London, 1954), Vol.~1, pp.\ 103--179.

\bibitem{Delahay:65}
P. Delahay, {\em Double layer and electrode kinetics} (Wiley\--Interscience,
  New York, 1965).

\bibitem{Delahay:66}
P. Delahay, J. Phys. Chem. {\bf 70},  2373  (1966).

\bibitem{Newman:65}
J. Newman, Trans. Faraday Soc. {\bf 61},  2229  (1965).

\bibitem{Newman:66}
W.-H. Smyrl and J. Newman, Trans. Faraday Soc. {\bf 62},  207  (1966).

\bibitem{Levie:67}
R.~D. Levie, Adv. Electrochem. Electrochem. Eng. {\bf 6},  380  (1967).

\bibitem{MacDonald:70}
J.~R. MacDonald, Trans. Farad. Soc. {\bf 66},  943  (1970).

\bibitem{Norton:90}
J.-D. Norton, H.-S. White, and S.-W. Feldberg, J. Phys. Chem. {\bf 94},  6772
  (1990).

\bibitem{Levine:71}
S. Levine, J. Colloid Int. Sciences {\bf 3},  619  (1971).

\bibitem{Fawcett:72}
W. Fawcett, Electroanal. Chem. and Int. Electrochem. {\bf 39},  474  (1972).

\bibitem{Damaskin:93}
B. Damaskin, V. Safonov, and N. Federovich, J. Electroanal. Chem. {\bf 349},  1
   (1993).

\bibitem{Damaskin:93b}
B. Damaskin, V. Safonov, and N. Federovich, Elektrokhymiya (translation) {\bf
  29},  1124  (1993).

\bibitem{Fawcett:98}
W. Fawcett,  in {\em Double layer effects in the electrode kinetics of electron
  and ion transfer reactions}, edited by J. Lipkowski and P. Ross (Wiley-VCH,
  N.Y., 1998), Chap.~8, pp.\ 323--371.

\bibitem{Fawcett:73}
W. Fawcett, Electroanal. Chem. and Int. Electrochem. {\bf 43},  175  (1973).

\bibitem{Levich:49}
B. Levich, Doklady Akad. Nauk SSSR {\bf 67},  309  (1949).

\bibitem{Gierst:61}
L. Gierst,  in {\em Transactions of the Symposium on Electrode Processes}, {\em
  Electrochemical Society Series}, edited by Yeager (Wiley, N.Y., 1961), p.\
  109.

\bibitem{Borukhov:97}
I. Borukhov, D. Andelman, and H. Orland, Phys. Rev. Lett. {\bf 79},  435
  (1997).

\bibitem{Bazant:00}
M. Bazant, asymptotic analysis of diffuse charge and limiting current in binary
  electrochemical cells (unpublished).

\bibitem{Brumleve:78}
T. Brumeleve and R. Buck, J. Electroanal. Chem {\bf 90},  1  (1978).

\bibitem{Murphy:92}
W.-D. Murphy, J.-A. Manzanares, S. Maf\'e, and H. Reiss, J. Phys. Chem. {\bf
  96},  9983  (1992).

\bibitem{Grafov:62}
B.~M. Grafov and A.~A. Chernenko, Doklady Akad. Nauk S.S.S.R. {\bf 146},  135
  (1962).

\bibitem{Chernenko:63}
A.~A. Chernenko, Doklady Akad. Nauk S.S.S.R. {\bf 153},  1963  (1963).

\bibitem{Rubinstein:79}
I. Rubinstein and L. Shtilman, J. Chem. Soc. Faraday Trans. II {\bf 75},  231
  (1979).

\bibitem{Henry:89}
J. Henry and B. Luoro, Nonlinear Analysis {\bf 13},  787  (1989).

\bibitem{Koper:93b}
M. Koper and J. Sluyters, J. Electroanal. Chem {\bf 352},  51  (1993).

\bibitem{Koper:96}
M. Koper,  in {\em Advances in Chemical Physics}, edited by I. Prigogine and S.
  Rice (John Wiley and Sons, New York, 1996), Vol.~XCII, pp.\ 161--298.

\bibitem{Strasser:99}
P. Strasser, M. Eiswirth, and M.~T.-M. Koper, J. Electroanal. Chem {\bf 478},
  50  (1999).

\bibitem{Krischer:99}
K. Krischer,  in {\em Modern aspects of electrochemistry}, edited by J.~B.
  B.E.~Conway and R. White (Academic/Plenum Publishers, New York, 1999),
  Vol.~32, pp.\ 1--142.

\bibitem{Newman:91}
J. Newman, {\em Electrochemical Systems} (Prentice Hall, Englewood Cliff, New
  Jersey, 1991).

\bibitem{Guidelli:92}
{\em Electrified interfaces in physics, chemistry and biology}, Vol.~355 of
  {\em NATO ASI Series C: Mathematical and Physical Sciences}, edited by R.
  Guidelli (Kluwer Academic Publishers, Nertherlands, 1992).

\bibitem{Bender:78}
C.~M. Bender and S.~A. Orszag, {\em Mathematical Methods for Scientists and
  Engineers} (McGraw-Hill, New York, 1978).

\bibitem{MacDonald:74}
J.~R. MacDonald, Electroanalytical Chemistry and Interf. Electrochem. {\bf 53},
   1  (1974).

\bibitem{MacDonald:79}
J.~R. MacDonald, J. Electroanal. Chem. {\bf 99},  283  (1979).

\bibitem{MacDonald:87}
J.~R. MacDonald, J. Electroanal. Chem. {\bf 223},  1  (1987).

\bibitem{Bamford:86}
{\em Chemical kinetics}, edited by C. Bamford and R. Compton (Elsevier,
  Amsterdam, 1986), Vol.~26.

\bibitem{Oldham:92}
K. Oldham, J. Electroanal. Chem {\bf 337},  91  (1992).

\bibitem{Bento:98}
M. Bento, L. Thouin, and C. Amatore, J. Electroanal. Chem {\bf 446},  91
  (1998).

\bibitem{Bento:98b}
M. Bento, L. Thouin, C. Amatore, and M. Montenegro, J. Electroanal. Chem {\bf
  443},  137  (1998).

\bibitem{Amatore:99}
C. Amatore, L. Thouin, and M.-F. Bento, J. Electroanal. Chem {\bf 463},  45
  (1999).

\bibitem{Ajdari:00}
A. Ajdari, Phys. Rev. E {\bf 61},  R45  (2000).

\end{thebibliography}

\section{Appendix A. List of Symbols}

\[
\begin{array}{ll}
D & \mbox{diffusion coefficient} \\
\mu & \mbox{mobility} \\
z_C,z_A,z & \mbox{charge numbers} \; \; \; z_C = -z_A = z \\
F & \mbox{Faraday's constant} \\
R & \mbox{universal gas constant} \\
T & \mbox{absolute temperature} \\
t & \mbox{time} \\
K_O, K_R & \mbox{kinetic constants} \\
k_O, k_R & \mbox{dimensionless kinetic constants} \\
\Cs & \mbox{capacitance of the Stern layer} \\
\epsb & \mbox{permittivity of the bulk solvent} \\
\epss & \mbox{effective permittivity of the Stern layer} \\
\lambda_S & \mbox{effective width of the Stern layer} \\
\lambda_D & \mbox{Debye screening length} \\
L & \mbox{distance between the electrodes (working and counter)} \\
\delta & = \lambda_S/\lambda_D \\
\epsilon & =\lambda_D/L \\
J & \mbox{current density} \\
J_F & \mbox{Faradaic current} \\
J_{DL} & \mbox{diffusion-limited current} \\
J_{RL} & \mbox{reaction-limited current} \\
\Cc, \Ca, \Cn & \mbox{concentrations of the species C, A and N} \\
\Phi & \mbox{electric scalar potential} \\
C^* & \mbox{mean anion concentration} \\
c_C, c_A, c_N & \mbox{dimensionless concentrations}  \\
c & \mbox{dimensionless average concentration of charged species} \\
\rho & \mbox{dimensionless charge density} \\
\phi & \mbox{dimensionless electrical potential} \\
j & \mbox{dimensionless current density} \\
j_F & \mbox{dimensionless Faradaic current} \\
j_d & \mbox{dimensionless displacement current} \\
\dphii & \mbox{dimensionless double layer voltage}  \\
\dphid & \mbox{dimensionless diffuse layer voltage}  \\
\dphis & \mbox{dimensionless Stern layer voltage}  \\
\end{array} 
\]

\end{document}